\documentclass[a4paper]{jpconf}
\usepackage{graphicx}
\usepackage{hyperref}

\usepackage{slashed}
\usepackage{xspace}
\usepackage{color}
\usepackage{amsmath}
\usepackage{amssymb}
\usepackage{mathtools}
\usepackage{subfig}
\usepackage[square,numbers,sort&compress]{natbib}

\usepackage{float}
\floatstyle{plaintop}
\restylefloat{table}

\graphicspath{{figures/}{jetdistr/}}

\newcommand{\NGluon}{{\sc NGluon}\xspace}
\newcommand{\NJet}{{\sc NJet}\xspace}

\def\HThat{\ensuremath{{\widehat{H}_T}}}
\def\R#1{\ensuremath{{\cal R}_{#1}}}
\def\as{\ensuremath{\alpha_s}}
\def\mur{\ensuremath{\mu_r}}
\def\muf{\ensuremath{\mu_f}}

\def\NLO{\mbox{\scriptsize NLO}}
\def\njet{n\text{-jet}}
\def\parton{{\mbox{\scriptsize parton}}}

\def\Fig#1{Fig.~\ref{#1}}
\def\Tab#1{Tab.~\ref{#1}}

\def\GeV{\text{ GeV}}

\begin{document}
\setlength{\textfloatsep}{20pt}
\title{Computation of multi-leg amplitudes with \NJet}

\author{S Badger$^1$, B Biedermann$^2$, P Uwer$^3$ and V Yundin$^{4,}$%
\footnote[5]{Speaker; talk given at the Workshop on Advanced Computing and Analysis Techniques in Physics (ACAT), Beijing, China, May 2013.%
}}
\address{%
$^1$ Theory Division, Physics Department, CERN, CH-1211 Geneva 23, Switzerland}
\address{%
$^2$ Universit\"at W\"urzburg, Institut f\"ur Theoretische Physik und Astrophysik,\\%
\hspace{1ex} Emil-Hilb-Weg 22, 97074 W\"urzburg, Germany}
\address{%
$^3$ Humboldt-Universit\"at zu Berlin, Institut f\"ur Physik,\\%
\hspace{1ex} Newtonstra{\ss}e 15, 12489 Berlin, Germany}
\address{%
$^4$ Max-Planck-Institut f\"ur Physik, F\"ohringer Ring 6, 80805 Munich, Germany}
\ead{yundin@mpp.mpg.de}

\begin{abstract}
In these proceedings we report our progress in the development of the publicly available C++ library \NJet for accurate calculations of high-multiplicity one-loop amplitudes.
As a phenomenological application we present the first complete next-to-leading order (NLO) calculation of five jet cross section at hadron colliders.
\end{abstract}

\section{Introduction}

NLO predictions of multi-jet production at hadron colliders have a long
history.  They are important processes for the LHC both as precision tests of
QCD and direct probes of the strong coupling and also as background in many new
physics searches.  The LHC experiments have been able to measure jet rates
for up to 6 hard jets which are now being used in new physics
searches \cite{Aad:2011tqa,Chatrchyan:2013gia,Chatrchyan:2013iqa}.  This
presents a serious challenge for precise theoretical predictions since high
multiplicity computations in QCD are notoriously difficult. Di-jet
production has been known at NLO for more than 20 years~\cite{Giele:1993dj} and has recently
seen improvements via NLO plus parton shower (NLO+PS) description~\cite{Alioli:2010xa,Hoeche:2012fm}
and steady progress towards NNLO QCD results~\cite{Ridder:2013mf}.
The full three-jet computation was completed and implemented in a public code {\sc
NLOJET++} 10 years ago~\cite{Nagy:2001fj}. Recently predictions for four-jet
production have been presented by two independent
groups~\cite{Bern:2011ep,Badger:2012pf}.

The advances in methods of evaluation of multi-leg virtual
amplitudes~\cite{Bern:1994zx,Bern:1994cg,%
Cascioli:2011va,Becker:2011vg,Actis:2012qn,Mastrolia:2010nb,Britto:2004nc,Ellis:2007br,Forde:2007mi,Giele:2008ve,%
Badger:2008cm,Ossola:2006us,Mastrolia:2012an,Mastrolia:2012bu,vanHameren:2009vq}
have inspired many efforts to automate NLO computations~%
\cite{Badger:2010nx,Hirschi:2011pa,Berger:2008sj,Bevilacqua:2011xh,Cullen:2011ac,Badger:2012pg}.
Processes with four final states, previously out of reach, can now be
routinely used for phenomenological
predictions~\cite{Bevilacqua:2012em,Greiner:2012im,Bern:2013gka,Cullen:2013saa,vanDeurzen:2013xla,Gehrmann:2013bga,Campanario:2013fsa}.
We refer the reader to other contributions to these proceedings for further details on the current
state-of-the art~\cite{Ossola:2013jea,Cullen:2013cka,Bern:2013pya}.

Five partons in the final state still constitute a considerable challenge, though steady progress in that direction
gives hope for the same level of automation in the near future.
Recent state-of-the-art calculations with five QCD partons in the final state include
the NLO QCD corrections to $pp\to W+5j$~\cite{Bern:2013gka} by the {\sc BlackHat} collaboration
 and NLO QCD corrections to $pp\to 5j$~\cite{Badger:2013yda}.

\section{5-Jet production at the LHC at 7 and 8~TeV \label{sec:5j}}

The different parts of the calculation, which contribute to NLO cross section can be schematically written as
\begin{gather}
  \delta\sigma^{\NLO} = \int\limits_n
  \big(d\sigma_n^{\rm V}
  + \int\limits_1 d\sigma_{n+1}^{\rm S}\big)
  + \int\limits_n d\sigma_n^{\rm Fac}
  + \int\limits_{n+1} \big(d\sigma_{n+1}^{\rm R} - d\sigma_{n+1}^{\rm S}\big).
\end{gather}
We used the Sherpa Monte-Carlo event generator \cite{Gleisberg:2008ta} to handle phase-space integration and
generation of tree-level amplitudes and Catani-Seymour dipole subtraction terms as implemented in Comix~\cite{Gleisberg:2008fv,Gleisberg:2007md}.

The one-loop matrix elements for the virtual corrections $d\sigma_n^{\rm V}$ are
evaluated with the publicly available \NJet\footnote{To download \NJet visit the project home page
at\\ \url{https://bitbucket.org/njet/njet/}.} package~\cite{Badger:2012pg}
interfaced to Sherpa via the Binoth Les Houches Accord \cite{Binoth:2010xt,Alioli:2013nda}.
\NJet is based on the \NGluon library~\cite{Badger:2010nx} %
and uses an on-shell generalized unitarity framework \cite{Britto:2004nc,Ellis:2007br,Forde:2007mi,Giele:2008ve,Badger:2008cm,Ossola:2006us}
 to compute multi-parton one-loop primitive amplitudes from tree-level building blocks~\cite{Berends:1987me}.
The scalar loop integrals are obtained via the {\sc QCDLoop/FF} package~\cite{vanOldenborgh:1990yc,Ellis:2007qk}.

\NJet implements full-colour expressions for up-to five outgoing QCD partons.
The complexity of high-multiplicity virtual corrections motivates us to explore
 ways to speed up the computation. One of the optimizations implemented in \NJet is the usage
 of de-symmetrized colour sums for multi-gluon final states, which allows us to get full colour result
 at a small fraction of computational cost by exploiting the Bose symmetry of the phase space~\cite{Ellis:2009zw,Badger:2012pg}.
Another possibility is to separate leading and sub-leading contributions,
which enables Monte-Carlo integrator to sample the dominant however simpler terms more often
and get the same statistical error with fewer evaluations of the expensive sub-leading part.

\begin{figure}[h]
\centering
    \includegraphics[width=0.43\textwidth]{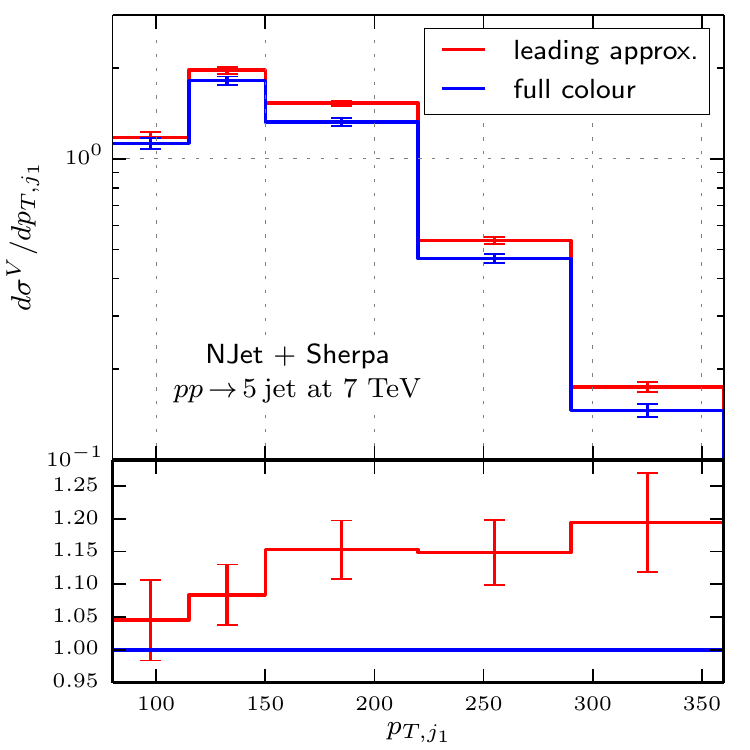}
  \caption{Full colour and leading approximation (as explained in the text)
  for the virtual corrections to the transverse momentum of the 1st jet in $pp\to 5j$.}
  \label{fig:colour}
\end{figure}

In our leading terms we include all multi-quark processes in the
large $N_c$ limit and processes with two or more gluons in the final state
using the de-symmetrized colour sums.
In Figure~\ref{fig:colour} we
compare leading and full virtual contributions to the hadrest jet transverse momentum in $pp\to 5j$.
The correction from the sub-leading part is around $10\%$ at low $p_T$ and shows a tendency to
grow with increasing hardness of the jet.
Considering that $d\sigma_n^{\rm V}$ contribute $\sim 50\%$ of the total NLO cross section for this process
this translates to $5-10$ percent effect depending on the kinematic region.

The calculation is done in QCD with five massless quark flavours including the bottom-quark in the
initial state. We neglect contributions from top quark loops. We set the renormalization scale equal
 to the factorization scale ($\mur=\muf=\mu$) and
use a dynamical scale based on the total transverse momentum $\HThat$ of the final state partons:
\begin{equation}
  \HThat = \sum_{i=1}^{N_\parton} p_{T,i}^\parton.
\end{equation}

For the definition of physical observables we use the anti-kt jet clustering algorithm as implemented
in {\sc FastJet} \cite{Cacciari:2011ma,Cacciari:2008gp}. We apply asymmetric cuts on the jets ordered
in transverse momenta, $p_T$, to match the ATLAS multi-jet measurements \cite{Aad:2011tqa}:
\begin{align}
  p_T^{j_1} &> 80 \GeV & p_T^{j_{\geq 2}} &> 60 \GeV & R &= 0.4
  \label{eq:cuts}
\end{align}

The PDFs are obtained through the LHAPDF interface \cite{Whalley:2005nh} with all central values using
NNPDF2.1~\cite{Ball:2011uy} for LO ($\alpha_s(M_Z) = 0.119$) and NNPDF2.3~\cite{Ball:2012cxX} for NLO
($\alpha_s(M_Z) = 0.118$) if not mentioned otherwise.

Generated events are stored in ROOT Ntuple format~\cite{Binoth:2010ra} which allows for flexible
analysis.  Renormalization and factorization scales can be changed at the analysis level as
well as the PDF set. This technique makes it possible to do extended analysis of
PDF uncertainties and scale dependence, which would otherwise be prohibitively expensive for such
high multiplicity processes.

\subsection{Numerical results \label{sec:results}}

\begin{figure}[htbp]
\centering
    \leavevmode
\subfloat[]{\label{fig:5j_scalevar}%
    \includegraphics[width=0.43\columnwidth]{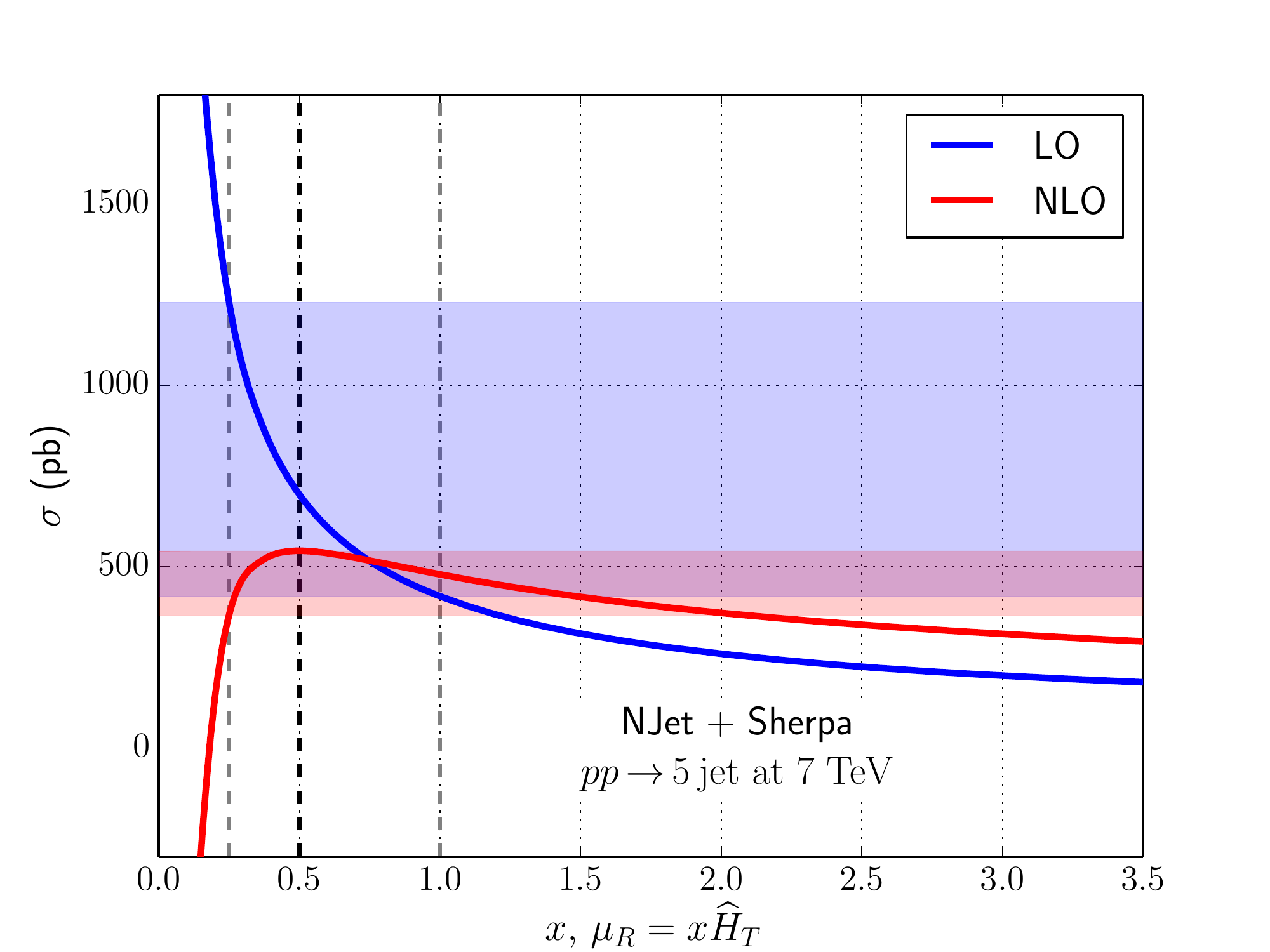}}
\subfloat[]{\label{fig:5j_scalevar_nlopdfs}%
    \includegraphics[width=0.43\columnwidth]{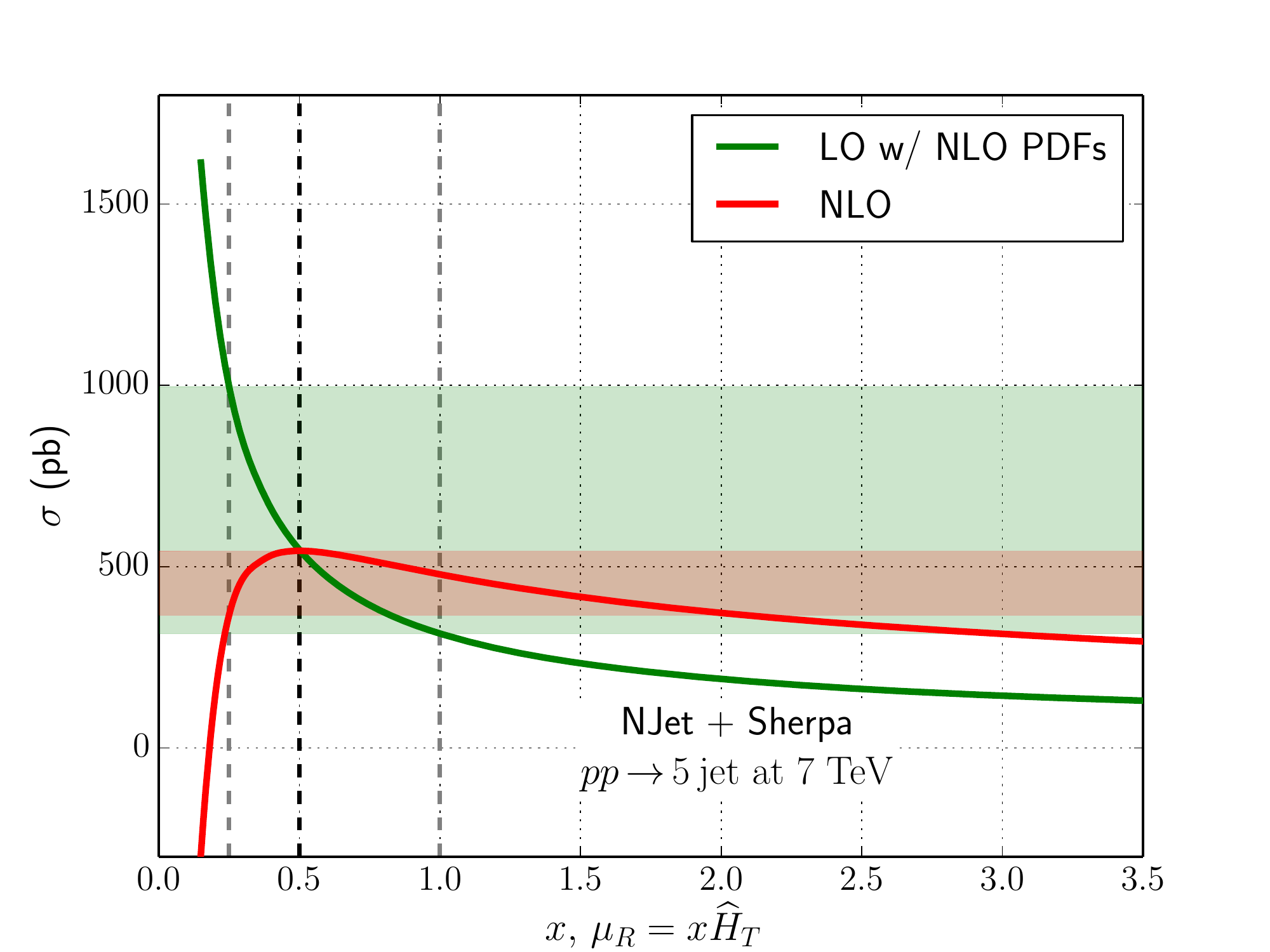}}
    \caption{Residual scale dependence of the 5-jet cross section in
    leading and next-to-leading order using LO~(a) and NLO~(b) PDFs for LO prediction.}
    \label{fig:5j_scalevar_all}
\end{figure}

Using the above setup we obtain for the 5-jet cross section at 7~TeV
\begin{align}
  \sigma_5^{\text{7TeV-LO}}(\mu=\HThat/2)  &= 0.699 ( 0.004 )^{+ 0.530 }_{- 0.280 }\: {\rm nb}, \\
  \sigma_5^{\text{7TeV-NLO}}(\mu=\HThat/2) &= 0.544 ( 0.016 )^{+ 0.0 }_{- 0.177 }\: {\rm nb}.
  \label{eq:5jXS7TeV}
\end{align}
In parentheses we quote the uncertainty due to the numerical integration.
The theoretical uncertainty has been estimated from scale variations over the range
$\mu\in[\HThat/4,\HThat]$ and is indicated by the sub- and superscripts.
As seen in \Fig{fig:5j_scalevar_all} the total cross section at the scale $\mu = \HThat$ is lower than
the central value which is the origin of the zero value of the upper error bound. The total cross
section at this scale is $\sigma_5^{\text{7TeV-NLO}}(\mu=\HThat) = 0.544 (0.016)\: {\rm nb}$.
For a centre-of-mass energy of 8~TeV the results read:
\begin{align}
  \sigma_5^{\text{8TeV-LO}}(\mu=\HThat/2)  &= 1.044 ( 0.006 )^{+ 0.770 }_{- 0.413 }\: {\rm nb}, \\
  \sigma_5^{\text{8TeV-NLO}}(\mu=\HThat/2) &= 0.790 ( 0.021 )^{+ 0.0 }_{- 0.313 }\: {\rm nb},
  \label{eq:5jXS8TeV}
\end{align}
where we have found $\sigma_5^{\text{8TeV-NLO}}(\mu=\HThat) = 0.723 ( 0.011 )\: {\rm nb}$.

As usual for a next-to-leading order correction a significant reduction of the scale uncertainty can
be observed.
In \Fig{fig:5j_scalevar_all} the scale dependence of the LO and NLO
cross section is illustrated. The dashed black line indicates the central scale $\mu=\HThat/2$. The horizontal bands show
the cross section uncertainty estimated by a scale variation within $\mu\in[\HThat/4,\HThat$].
%  The uncertainty due to scale variation is roughly reduced by a factor of one third. Furthermore we see
% that around $\mu=\HThat/2$ the NLO cross section is flat indicating that $\mu=\HThat/2$ is a
% reasonable choice for the central scale.
% This is further supported by the fact that for
% $\mu=\HThat/2$ the NLO corrections are very small.

By comparing Figs.~\ref{fig:5j_scalevar} and \ref{fig:5j_scalevar_nlopdfs} we observe that a significant part of the NLO corrections comes
from using NLO PDFs with the corresponding $\as$. Similar to what has been found in Ref.~\cite{Badger:2012pf} we
conclude that using the NLO PDFs in the LO predictions gives a
better approximation to the full result compared to using LO PDFs.

In \Tab{tab:xs} we show for completeness the cross sections
for two, three and four-jet production as calculated with \NJet using
the same setup as in the five jet case.
\begin{table}[h]
\centering
  \setlength{\tabcolsep}{12pt}
  \renewcommand{\arraystretch}{1.5}
  \begin{tabular}{lcc}
    \hline
    $n$ & $\sigma_n^{\text{7TeV-LO}} \: [{\rm nb}]$ & $\sigma_n^{\text{7TeV-NLO}} \: [{\rm nb}]$ \\
    \hline
    $2$ & $768.0 ( 0.9 )^{+ 203.0}_{- 151.3 }$ & $1175 ( 3    )^{+ 120 }_{- 129  }$  \\
    \hline
    $3$ & $71.1 (  0.1 )^{+ 31.5 }_{- 20.0  }$ & $52.5 ( 0.3  )^{+ 1.9 }_{- 19.3 }$  \\
    \hline
    $4$ & $7.23 ( 0.02 )^{+ 4.37 }_{- 2.50  }$ & $5.65 ( 0.07 )^{+ 0   }_{- 1.93 }$ \\
    \hline
  \end{tabular}
  \caption{Cross sections for 2, 3 and 4 jets at 7~TeV.}
  \label{tab:xs}
\end{table}

The jet rates have been measured
recently by ATLAS using the 7~TeV data set~\cite{Aad:2011tqa}.
\begin{figure}[htbp]
\centering
    \leavevmode
\subfloat[]{\label{fig:MJincl}%
    \includegraphics[width=0.43\columnwidth]{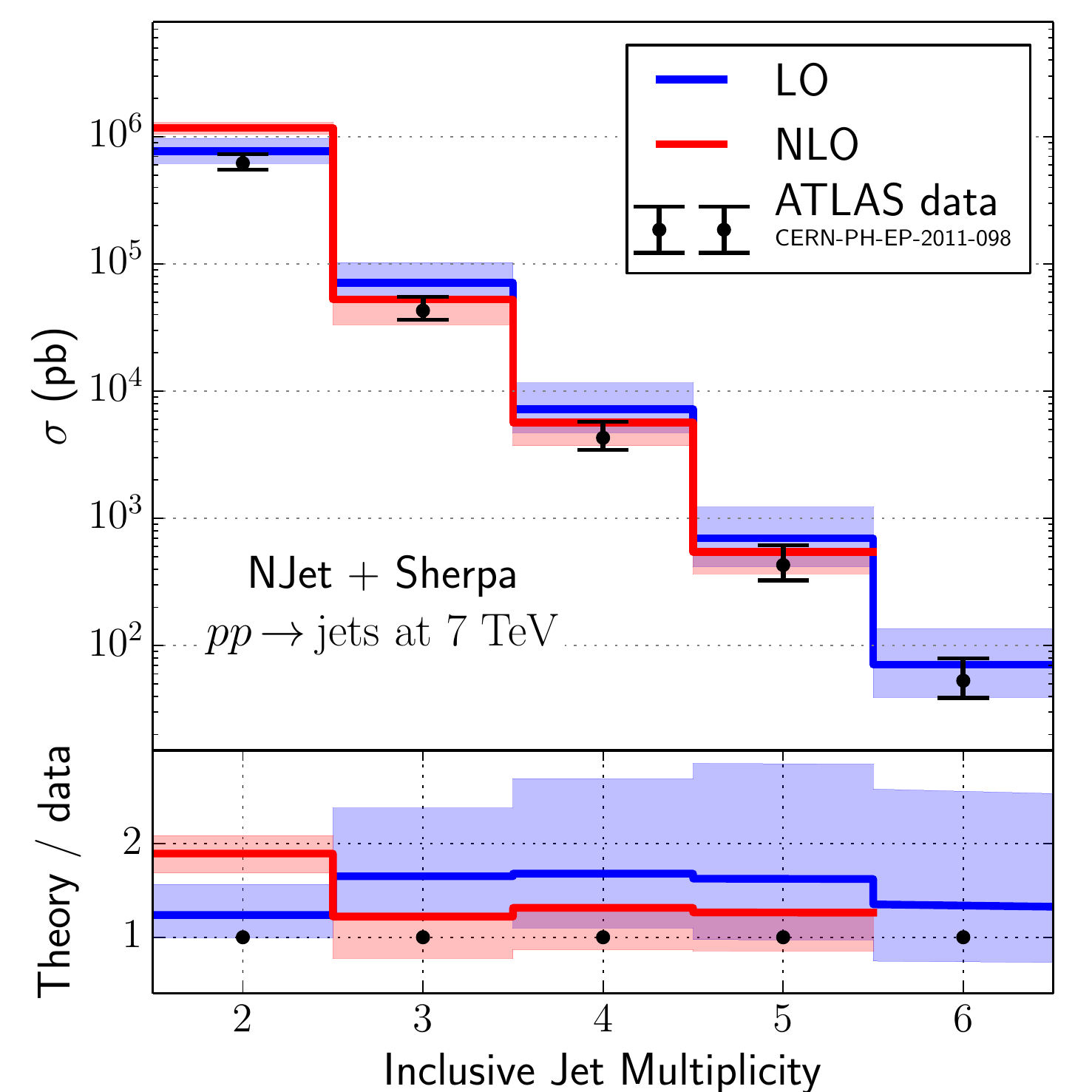}}
\subfloat[]{\label{fig:jetratios}%
    \includegraphics[width=0.43\columnwidth]{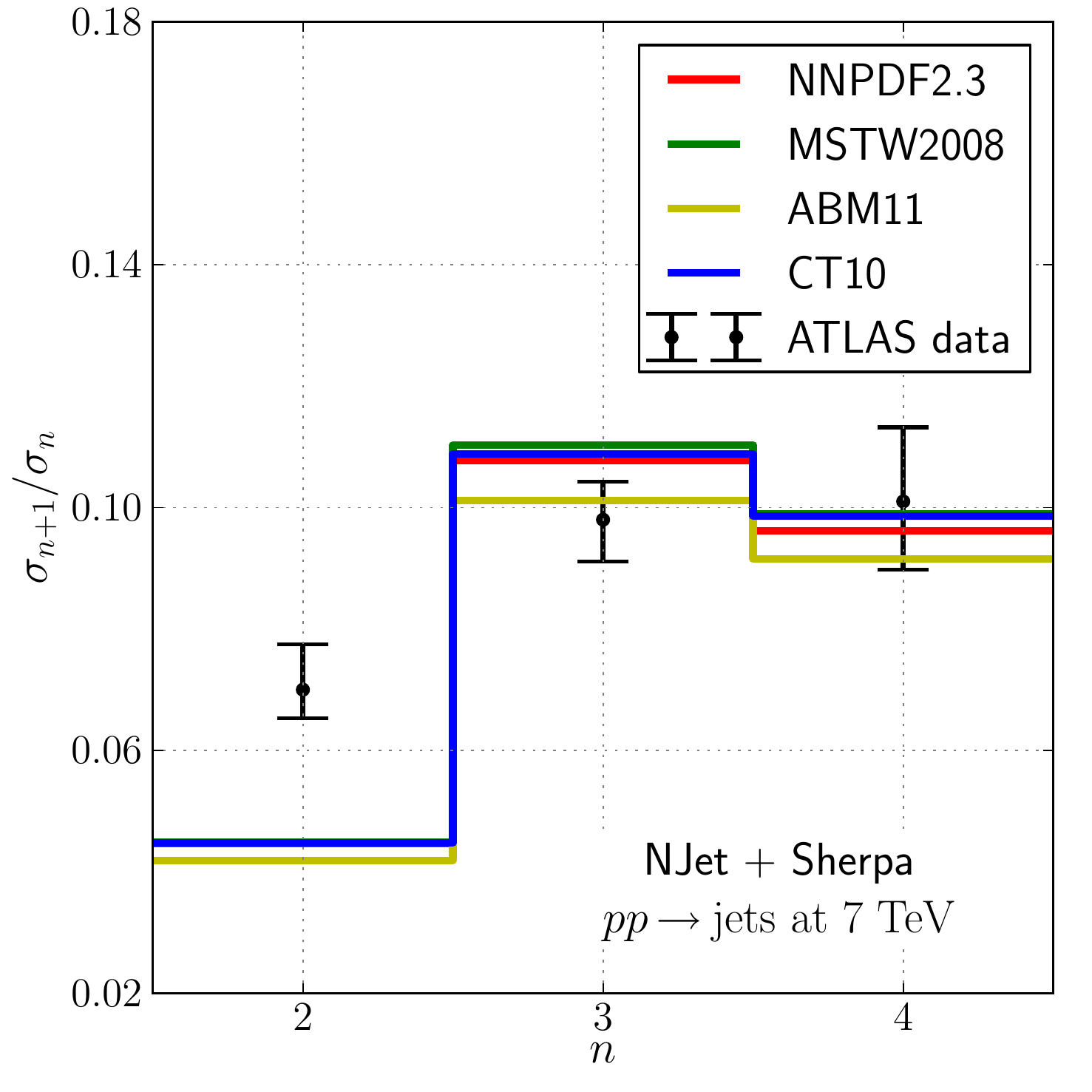}}
    \caption{(a) LO and NLO cross sections for jet production calculated with \NJet as well
    as results from ATLAS measurements \cite{Aad:2011tqa}.
    (b) NLO \NJet predictions with different PDF sets for the jet ratios ${\cal{R}}_n$ compared
    with recent ATLAS measurements \cite{Aad:2011tqa}.
    }
\end{figure}
In \Fig{fig:MJincl} we show the data together with the theoretical predictions in leading and
next-to-leading order. In case of the six jet rate only LO results are shown. In the lower plot the
ratio of theoretical predictions with respect to data is given. With exception of the two-jet cross
section the inclusion of the NLO results improves significantly the agreement with data.

\begin{figure}[h]
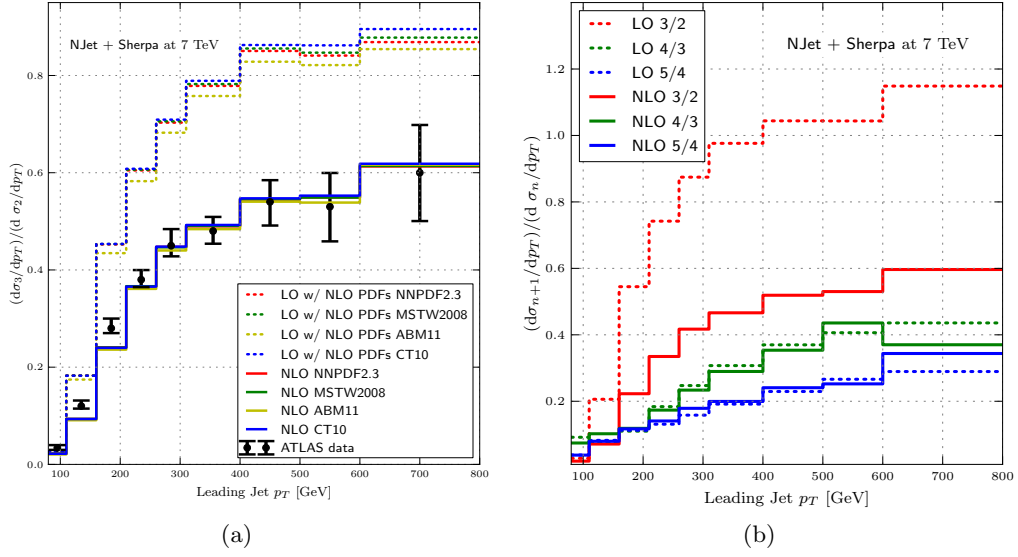

\centering
\subfloat[]{\label{fig:32ratiodata}%
    \includegraphics[width=0.43\columnwidth]{{{plot_32ratio_data_pT_1}}}}
\subfloat[]{\label{fig:jetratio_pT}%
    \includegraphics[width=0.43\columnwidth]{{{plot_jetratio_pT_1}}}}
  \caption{(a) The 3/2 jet ratio as a function of the $p_T$ of the leading jet compared with ATLAS
  data~\cite{Aad:2011tqa} ($R=0.6$).
  (b) The $\R{n}$ ratios as functions of the $p_T$ of the leading jet ($R=0.4$).}
\end{figure}

In addition to inclusive cross
sections it is useful to consider their ratios since many theoretical
and experimental uncertainties may cancel between numerator
and denominator. In particular we consider
\begin{equation}
  \R{n} =  {\sigma_{(n+1)\text{-jet}}\over\sigma_{\njet}}.
\end{equation}
This quantity is in leading order proportional to the QCD coupling $\as$ and can be used to
determine the value of $\as$ from jet rates.
In \Fig{fig:jetratios} we show QCD predictions in NLO using different PDF sets together with the
results from ATLAS. The results obtained from NNPDF2.3 are also collected in \Tab{tab:jetratios}
where, in addition, the ratios at leading order (using the LO setup with NNPDF2.1) are shown.
\begin{table}[htbp]
  \setlength{\tabcolsep}{12pt}
  \renewcommand{\arraystretch}{1.6}
\centering
%    \leavevmode
    \begin{tabular}{cccc}
      \hline
      \R{n} & ATLAS~\cite{Aad:2011tqa} & LO & NLO \\ \hline
      2 & $0.070^{+ 0.007 }_{- 0.005 }$ & $0.0925(0.0002)$  & $0.0447(0.0003)$\\ \hline
      3 & $0.098^{+ 0.006 }_{- 0.007 }$ & $0.102(0.000)$  & $0.108(0.002)$\\ \hline
      4 & $0.101^{+ 0.012 }_{- 0.011 }$ & $0.097(0.001)$  & $0.096(0.003)$\\ \hline
      5 & $0.123^{+ 0.028 }_{- 0.027 }$ & $0.102(0.001)$  & $--$\\ \hline
    \end{tabular}
    \caption{Results for the jet ratios $\R{n}$ for the central scale of $\HThat/2$ and NNPDF2.3 PDF
    set.}
    \label{tab:jetratios}
\end{table}
In case of \R{3}\ and \R{4}\ perturbation theory seems to provide stable results. The leading order
and next-to-leading order values differ by less than 10\%. In addition NNPDF~\cite{Ball:2012cxX},
CT10~\cite{Lai:2010vv} and MSTW08~\cite{Martin:2009iq} give compatible predictions.
ABM11~\cite{Alekhin:2012ig} gives slightly smaller results for \R{3}\ and \R{4}.
Within uncertainties the predictions also agree with the ATLAS measurements.
The poor description of \R{2}\ can be attributed to the inclusive two-jet cross section
which seems to be inadequately described by a fixed order NLO calculation.
As a function of the leading jet $p_T$, all PDF
sets agree well with the 3/2 ratio ATLAS data at large $p_T$ as shown in \Fig{fig:32ratiodata}.
In \Fig{fig:jetratio_pT} we compare LO and NLO predictions for \R{n}\ as function of the leading jet
$p_T$. While for \R{3}\ and \R{4}\ the corrections are moderate for all values of $p_T$ we observe
large negative corrections independent from $p_T$ in case of \R{2}.

\begin{figure}[htbp]
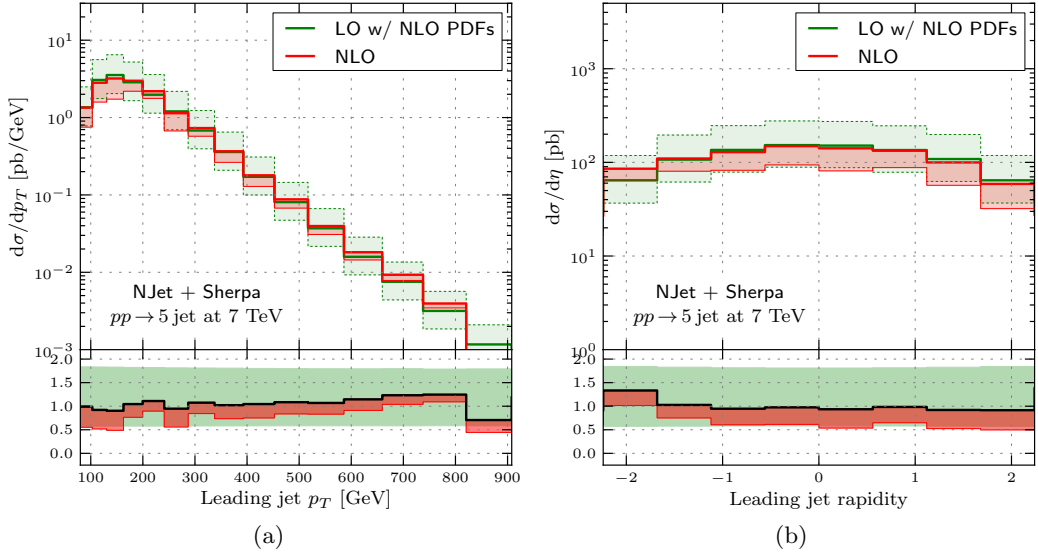

\centering
    \leavevmode
\subfloat[]{%
    \includegraphics[width=0.43\columnwidth]{{{plot_q16.4_l8_sm0.1_NNPDF23_7_nlopdf_jet_pT_1}}}}
\subfloat[]{%
    \includegraphics[width=0.43\columnwidth]{{{plot_q20.4_l10_sm0.1_NNPDF23_7_nlopdf_jet_eta_1}}}}
    \caption{The $p_T$ and rapidity distributions of the leading jet. Both LO and NLO use the NNPDF2.3 PDF set
    with $\alpha_s(M_Z) = 0.118$}
    \label{fig:7TeVpt1dist}
\end{figure}
In \Fig{fig:7TeVpt1dist} we show the transverse momentum and rapidity distributions
of the leading jet for five-jet production. Similarly to total cross section
we observe significant reduction of the scale uncertainty when going from LO to NLO.
 Using again the NLO setup to
calculate the LO predictions, the NLO calculation gives very small
corrections. Over a wide range the LO predictions are modified by less
than 10\%. A remarkable feature observed already in the 4-jet
calculation \cite{Bern:2011ep,Badger:2012pf} is the almost constant
K-factor.

\section{Conclusions}

In this contribution we have presented first results for five-jet production at NLO accuracy in QCD.
We find moderate corrections of the order of 10\% at NLO with respect to a leading order computation using NLO PDFs.
We have compared theoretical predictions for inclusive jet cross sections and jet
rates with data from ATLAS. With the exception of quantities affected by the two-jet rate we find
good agreement between theory and data.

\section*{Acknowledgements}

This work has been supported by the Helmholtz Gemeinschaft under contract HA-101 (Alliance Physics at the
Terascale), by the German Research Foundation (DFG) through the transregional collaborative research
centre ``Computational Particle Physics'' (SFB-TR9), by the European Commission through contract
PITN-GA-2010-264564 (LHCPhenoNet) and by the Alexander von Humboldt Foundation,
in the framework of the Sofja Kovaleskaja Award 2010, endowed by the German Federal Ministry of Education and Research.

\small
\bibliographystyle{iopart-num}
\bibliography{acatproc}

\providecommand{\newblock}{}
\begin{thebibliography}{10}
\expandafter\ifx\csname url\endcsname\relax
  \def\url#1{{\tt #1}}\fi
\expandafter\ifx\csname urlprefix\endcsname\relax\def\urlprefix{URL }\fi
\providecommand{\eprint}[2][]{\url{#2}}

\bibitem{Aad:2011tqa}
Aad G {\em et~al.\/} (ATLAS Collaboration) 2011 {\em Eur.Phys.J.\/} {\bf C71}
  1763 (\textit{Preprint} \href{http://xxx.lanl.gov/abs/1107.2092}{{\tt
  arXiv:1107.2092}})

\bibitem{Chatrchyan:2013gia}
Chatrchyan S {\em et~al.\/} (CMS Collaboration) 2013  (\textit{Preprint}
  \href{http://xxx.lanl.gov/abs/1311.1799}{{\tt arXiv:1311.1799}})

\bibitem{Chatrchyan:2013iqa}
Chatrchyan S {\em et~al.\/} (CMS Collaboration) 2013  (\textit{Preprint}
  \href{http://xxx.lanl.gov/abs/1311.4937}{{\tt arXiv:1311.4937}})

\bibitem{Giele:1993dj}
Giele W, Glover E~N and Kosower D~A 1993 {\em Nucl.Phys.\/} {\bf B403} 633--670
  (\textit{Preprint} \href{http://xxx.lanl.gov/abs/hep-ph/9302225}{{\tt
  hep-ph/9302225}})

\bibitem{Alioli:2010xa}
Alioli S, Hamilton K, Nason P, Oleari C and Re E 2011 {\em JHEP\/} {\bf 1104}
  081 (\textit{Preprint} \href{http://xxx.lanl.gov/abs/1012.3380}{{\tt
  arXiv:1012.3380}})

\bibitem{Hoeche:2012fm}
Hoeche S and Schonherr M 2012 {\em Phys.Rev.\/} {\bf D86} 094042
  (\textit{Preprint} \href{http://xxx.lanl.gov/abs/1208.2815}{{\tt
  arXiv:1208.2815}})

\bibitem{Ridder:2013mf}
Gehrmann-De~Ridder A, Gehrmann T, Glover E and Pires J 2013 {\em
  Phys.Rev.Lett.\/} {\bf 110} 162003 (\textit{Preprint}
  \href{http://xxx.lanl.gov/abs/1301.7310}{{\tt arXiv:1301.7310}})

\bibitem{Nagy:2001fj}
Nagy Z 2002 {\em Phys.Rev.Lett.\/} {\bf 88} 122003 (\textit{Preprint}
  \href{http://xxx.lanl.gov/abs/hep-ph/0110315}{{\tt hep-ph/0110315}})

\bibitem{Bern:2011ep}
Bern Z, Diana G, Dixon L, Febres~Cordero F, Hoeche S {\em et~al.\/} 2012 {\em
  Phys.Rev.Lett.\/} {\bf 109} 042001 (\textit{Preprint}
  \href{http://xxx.lanl.gov/abs/1112.3940}{{\tt arXiv:1112.3940}})

\bibitem{Badger:2012pf}
Badger S, Biedermann B, Uwer P and Yundin V 2013 {\em Phys.Lett.\/} {\bf B718}
  965--978 (\textit{Preprint} \href{http://xxx.lanl.gov/abs/1209.0098}{{\tt
  arXiv:1209.0098}})

\bibitem{Bern:1994zx}
Bern Z, Dixon L~J, Dunbar D~C and Kosower D~A 1994 {\em Nucl.Phys.\/} {\bf
  B425} 217--260 (\textit{Preprint}
  \href{http://xxx.lanl.gov/abs/hep-ph/9403226}{{\tt hep-ph/9403226}})

\bibitem{Bern:1994cg}
Bern Z, Dixon L~J, Dunbar D~C and Kosower D~A 1995 {\em Nucl.Phys.\/} {\bf
  B435} 59--101 (\textit{Preprint}
  \href{http://xxx.lanl.gov/abs/hep-ph/9409265}{{\tt hep-ph/9409265}})

\bibitem{Cascioli:2011va}
Cascioli F, Maierhofer P and Pozzorini S 2012 {\em Phys.Rev.Lett.\/} {\bf 108}
  111601 (\textit{Preprint} \href{http://xxx.lanl.gov/abs/1111.5206}{{\tt
  arXiv:1111.5206}})

\bibitem{Becker:2011vg}
Becker S, Goetz D, Reuschle C, Schwan C and Weinzierl S 2012 {\em
  Phys.Rev.Lett.\/} {\bf 108} 032005 (\textit{Preprint}
  \href{http://xxx.lanl.gov/abs/1111.1733}{{\tt arXiv:1111.1733}})

\bibitem{Actis:2012qn}
Actis S, Denner A, Hofer L, Scharf A and Uccirati S 2013 {\em JHEP\/} {\bf
  1304} 037 (\textit{Preprint} \href{http://xxx.lanl.gov/abs/1211.6316}{{\tt
  arXiv:1211.6316}})

\bibitem{Mastrolia:2010nb}
Mastrolia P, Ossola G, Reiter T and Tramontano F 2010 {\em JHEP\/} {\bf 1008}
  080 (\textit{Preprint} \href{http://xxx.lanl.gov/abs/1006.0710}{{\tt
  arXiv:1006.0710}})

\bibitem{Britto:2004nc}
Britto R, Cachazo F and Feng B 2005 {\em Nucl.Phys.\/} {\bf B725} 275--305
  (\textit{Preprint} \href{http://xxx.lanl.gov/abs/hep-th/0412103}{{\tt
  hep-th/0412103}})

\bibitem{Ellis:2007br}
Ellis R~K, Giele W and Kunszt Z 2008 {\em JHEP\/} {\bf 0803} 003
  (\textit{Preprint} \href{http://xxx.lanl.gov/abs/0708.2398}{{\tt
  arXiv:0708.2398}})

\bibitem{Forde:2007mi}
Forde D 2007 {\em Phys.Rev.\/} {\bf D75} 125019 (\textit{Preprint}
  \href{http://xxx.lanl.gov/abs/0704.1835}{{\tt arXiv:0704.1835}})

\bibitem{Giele:2008ve}
Giele W~T, Kunszt Z and Melnikov K 2008 {\em JHEP\/} {\bf 0804} 049
  (\textit{Preprint} \href{http://xxx.lanl.gov/abs/0801.2237}{{\tt
  arXiv:0801.2237}})

\bibitem{Badger:2008cm}
Badger S 2009 {\em JHEP\/} {\bf 0901} 049 (\textit{Preprint}
  \href{http://xxx.lanl.gov/abs/0806.4600}{{\tt arXiv:0806.4600}})

\bibitem{Ossola:2006us}
Ossola G, Papadopoulos C~G and Pittau R 2007 {\em Nucl.Phys.\/} {\bf B763}
  147--169 (\textit{Preprint}
  \href{http://xxx.lanl.gov/abs/hep-ph/0609007}{{\tt hep-ph/0609007}})

\bibitem{Mastrolia:2012an}
Mastrolia P, Mirabella E, Ossola G and Peraro T 2012 {\em Phys.Lett.\/} {\bf
  B718} 173--177 (\textit{Preprint}
  \href{http://xxx.lanl.gov/abs/1205.7087}{{\tt arXiv:1205.7087}})

\bibitem{Mastrolia:2012bu}
Mastrolia P, Mirabella E and Peraro T 2012 {\em JHEP\/} {\bf 1206} 095
  (\textit{Preprint} \href{http://xxx.lanl.gov/abs/1203.0291}{{\tt
  arXiv:1203.0291}})

\bibitem{vanHameren:2009vq}
van Hameren A 2009 {\em JHEP\/} {\bf 0907} 088 (\textit{Preprint}
  \href{http://xxx.lanl.gov/abs/0905.1005}{{\tt arXiv:0905.1005}})

\bibitem{Badger:2010nx}
Badger S, Biedermann B and Uwer P 2011 {\em Comput.Phys.Commun.\/} {\bf 182}
  1674--1692 (\textit{Preprint} \href{http://xxx.lanl.gov/abs/1011.2900}{{\tt
  arXiv:1011.2900}})

\bibitem{Hirschi:2011pa}
Hirschi V, Frederix R, Frixione S, Garzelli M~V, Maltoni F {\em et~al.\/} 2011
  {\em JHEP\/} {\bf 1105} 044 (\textit{Preprint}
  \href{http://xxx.lanl.gov/abs/1103.0621}{{\tt arXiv:1103.0621}})

\bibitem{Berger:2008sj}
Berger C, Bern Z, Dixon L, Febres~Cordero F, Forde D {\em et~al.\/} 2008 {\em
  Phys.Rev.\/} {\bf D78} 036003 (\textit{Preprint}
  \href{http://xxx.lanl.gov/abs/0803.4180}{{\tt arXiv:0803.4180}})

\bibitem{Bevilacqua:2011xh}
Bevilacqua G, Czakon M, Garzelli M, van Hameren A, Kardos A {\em et~al.\/} 2013
  {\em Comput.Phys.Commun.\/} {\bf 184} 986--997 (\textit{Preprint}
  \href{http://xxx.lanl.gov/abs/1110.1499}{{\tt arXiv:1110.1499}})

\bibitem{Cullen:2011ac}
Cullen G, Greiner N, Heinrich G, Luisoni G, Mastrolia P {\em et~al.\/} 2012
  {\em Eur.Phys.J.\/} {\bf C72} 1889 (\textit{Preprint}
  \href{http://xxx.lanl.gov/abs/1111.2034}{{\tt arXiv:1111.2034}})

\bibitem{Badger:2012pg}
Badger S, Biedermann B, Uwer P and Yundin V 2013 {\em Comput.Phys.Commun.\/}
  {\bf 184} 1981--1998 (\textit{Preprint}
  \href{http://xxx.lanl.gov/abs/1209.0100}{{\tt arXiv:1209.0100}})

\bibitem{Bevilacqua:2012em}
Bevilacqua G and Worek M 2012 {\em JHEP\/} {\bf 1207} 111 (\textit{Preprint}
  \href{http://xxx.lanl.gov/abs/1206.3064}{{\tt arXiv:1206.3064}})

\bibitem{Greiner:2012im}
Greiner N, Heinrich G, Mastrolia P, Ossola G, Reiter T {\em et~al.\/} 2012 {\em
  Phys.Lett.\/} {\bf B713} 277--283 (\textit{Preprint}
  \href{http://xxx.lanl.gov/abs/1202.6004}{{\tt arXiv:1202.6004}})

\bibitem{Bern:2013gka}
Bern Z, Dixon L, Febres~Cordero F, Hoeche S, Ita H {\em et~al.\/} 2013 {\em
  Phys.Rev.\/} {\bf D88} 014025 (\textit{Preprint}
  \href{http://xxx.lanl.gov/abs/1304.1253}{{\tt arXiv:1304.1253}})

\bibitem{Cullen:2013saa}
Cullen G, van Deurzen H, Greiner N, Luisoni G, Mastrolia P {\em et~al.\/} 2013
  {\em Phys.Rev.Lett.\/} {\bf 111} 131801 (\textit{Preprint}
  \href{http://xxx.lanl.gov/abs/1307.4737}{{\tt arXiv:1307.4737}})

\bibitem{vanDeurzen:2013xla}
van Deurzen H, Luisoni G, Mastrolia P, Mirabella E, Ossola G {\em et~al.\/}
  2013 {\em Phys.Rev.Lett.\/} {\bf 111} 171801 (\textit{Preprint}
  \href{http://xxx.lanl.gov/abs/1307.8437}{{\tt arXiv:1307.8437}})

\bibitem{Gehrmann:2013bga}
Gehrmann T, Greiner N and Heinrich G 2013  (\textit{Preprint}
  \href{http://xxx.lanl.gov/abs/1308.3660}{{\tt arXiv:1308.3660}})

\bibitem{Campanario:2013fsa}
Campanario F, Figy T, Plätzer S and Sjödahl M 2013 {\em Phys.Rev.Lett.\/}
  {\bf 111} 211802 (\textit{Preprint}
  \href{http://xxx.lanl.gov/abs/1308.2932}{{\tt arXiv:1308.2932}})

\bibitem{Ossola:2013jea}
Ossola G 2013  (\textit{Preprint} \href{http://xxx.lanl.gov/abs/1310.3214}{{\tt
  arXiv:1310.3214}})

\bibitem{Cullen:2013cka}
Cullen G, van Deurzen H, Greiner N, Heinrich G, Luisoni G {\em et~al.\/} 2013
  (\textit{Preprint} \href{http://xxx.lanl.gov/abs/1309.3741}{{\tt
  arXiv:1309.3741}})

\bibitem{Bern:2013pya}
Bern Z, Dixon L, Cordero F~F, Hoeche S, Ita H {\em et~al.\/} 2013
  (\textit{Preprint} \href{http://xxx.lanl.gov/abs/1310.2808}{{\tt
  arXiv:1310.2808}})

\bibitem{Badger:2013yda}
Badger S, Biedermann B, Uwer P and Yundin V 2013  (\textit{Preprint}
  \href{http://xxx.lanl.gov/abs/1309.6585}{{\tt arXiv:1309.6585}})

\bibitem{Gleisberg:2008ta}
Gleisberg T, Hoeche S, Krauss F, Schonherr M, Schumann S {\em et~al.\/} 2009
  {\em JHEP\/} {\bf 0902} 007 (\textit{Preprint}
  \href{http://xxx.lanl.gov/abs/0811.4622}{{\tt arXiv:0811.4622}})

\bibitem{Gleisberg:2008fv}
Gleisberg T and Hoeche S 2008 {\em JHEP\/} {\bf 0812} 039 (\textit{Preprint}
  \href{http://xxx.lanl.gov/abs/0808.3674}{{\tt arXiv:0808.3674}})

\bibitem{Gleisberg:2007md}
Gleisberg T and Krauss F 2008 {\em Eur.Phys.J.\/} {\bf C53} 501--523
  (\textit{Preprint} \href{http://xxx.lanl.gov/abs/0709.2881}{{\tt
  arXiv:0709.2881}})

\bibitem{Binoth:2010xt}
Binoth T, Boudjema F, Dissertori G, Lazopoulos A, Denner A {\em et~al.\/} 2010
  {\em Comput.Phys.Commun.\/} {\bf 181} 1612--1622 (\textit{Preprint}
  \href{http://xxx.lanl.gov/abs/1001.1307}{{\tt arXiv:1001.1307}})

\bibitem{Alioli:2013nda}
Alioli S, Badger S, Bellm J, Biedermann B, Boudjema F {\em et~al.\/} 2013
  (\textit{Preprint} \href{http://xxx.lanl.gov/abs/1308.3462}{{\tt
  arXiv:1308.3462}})

\bibitem{Berends:1987me}
Berends F~A and Giele W 1988 {\em Nucl.Phys.\/} {\bf B306} 759

\bibitem{vanOldenborgh:1990yc}
van Oldenborgh G 1991 {\em Comput.Phys.Commun.\/} {\bf 66} 1--15

\bibitem{Ellis:2007qk}
Ellis R~K and Zanderighi G 2008 {\em JHEP\/} {\bf 0802} 002 (\textit{Preprint}
  \href{http://xxx.lanl.gov/abs/0712.1851}{{\tt arXiv:0712.1851}})

\bibitem{Ellis:2009zw}
Ellis R~K, Melnikov K and Zanderighi G 2009 {\em JHEP\/} {\bf 0904} 077
  (\textit{Preprint} \href{http://xxx.lanl.gov/abs/0901.4101}{{\tt
  arXiv:0901.4101}})

\bibitem{Cacciari:2011ma}
Cacciari M, Salam G~P and Soyez G 2012 {\em Eur.Phys.J.\/} {\bf C72} 1896
  (\textit{Preprint} \href{http://xxx.lanl.gov/abs/1111.6097}{{\tt
  arXiv:1111.6097}})

\bibitem{Cacciari:2008gp}
Cacciari M, Salam G~P and Soyez G 2008 {\em JHEP\/} {\bf 0804} 063
  (\textit{Preprint} \href{http://xxx.lanl.gov/abs/0802.1189}{{\tt
  arXiv:0802.1189}})

\bibitem{Whalley:2005nh}
Whalley M, Bourilkov D and Group R 2005  (\textit{Preprint}
  \href{http://xxx.lanl.gov/abs/hep-ph/0508110}{{\tt hep-ph/0508110}})

\bibitem{Ball:2011uy}
Ball R~D {\em et~al.\/} (NNPDF Collaboration) 2012 {\em Nucl.Phys.\/} {\bf
  B855} 153--221 (\textit{Preprint}
  \href{http://xxx.lanl.gov/abs/1107.2652}{{\tt arXiv:1107.2652}})

\bibitem{Ball:2012cxX}
Ball R~D, Bertone V, Carrazza S, Deans C~S, Del~Debbio L {\em et~al.\/} 2013
  {\em Nucl.Phys.\/} {\bf B867} 244--289 (\textit{Preprint}
  \href{http://xxx.lanl.gov/abs/1207.1303}{{\tt arXiv:1207.1303}})

\bibitem{Binoth:2010ra}
Andersen J {\em et~al.\/} (SM and NLO Multileg Working Group) 2010   21--189
  (\textit{Preprint} \href{http://xxx.lanl.gov/abs/1003.1241}{{\tt
  arXiv:1003.1241}})

\bibitem{Lai:2010vv}
Lai H~L, Guzzi M, Huston J, Li Z, Nadolsky P~M {\em et~al.\/} 2010 {\em
  Phys.Rev.\/} {\bf D82} 074024 (\textit{Preprint}
  \href{http://xxx.lanl.gov/abs/1007.2241}{{\tt arXiv:1007.2241}})

\bibitem{Martin:2009iq}
Martin A, Stirling W, Thorne R and Watt G 2009 {\em Eur.Phys.J.\/} {\bf C63}
  189--285 (\textit{Preprint} \href{http://xxx.lanl.gov/abs/0901.0002}{{\tt
  arXiv:0901.0002}})

\bibitem{Alekhin:2012ig}
Alekhin S, Blumlein J and Moch S 2012 {\em Phys.Rev.\/} {\bf D86} 054009
  (\textit{Preprint} \href{http://xxx.lanl.gov/abs/1202.2281}{{\tt
  arXiv:1202.2281}})

\end{thebibliography}

\end{document}